\documentstyle[12pt]{article}

\begin{document}
\frenchspacing 

\vspace*{1in}
\thispagestyle{empty}
\begin{center}

{\huge What is quantum field theory and\\[1ex]
 why have some physicists abandoned it?}
\bigskip

{\Large R. Jackiw \\[1ex]
 \large M\kern.1em I\kern.1em T }

\vfill

\centerline{\bf Abstract}
\medskip
\noindent
The present-day crisis in quantum field theory is described.

\vfill

{Contribution to  {\it Pulse},  the newsletter of the\\[1ex]Laboratory for
Nuclear Science}

\vfill

\centerline{MIT-CTP\#2671  \qquad  hep-th/9709212\qquad September 1997}

\end{center}

\newpage

Quantum field theory offers physicists a tremendously wide variety of applications: it
is both a language with which many physical processes can be discussed,
and also it provides a model for fundamental physics, the so-called ``standard model,"
which thus far has passed every experimental test.  No other framework exists in
which one can calculate so many phenomena with such ease and accuracy.

Arising from a mathematical account of the propagation of fluids (both
``ponderable'' and ``imponderable"), field theory emerged over a hundred years
ago in the description within classical physics of electromagnetism and soon
thereafter of gravity.  Schr\"{o}dinger's wave mechanics became a bridge between
classical and quantum field theory:  the quantum mechanical wave function is also a
local field, which when ``second" quantized gave rise to a true quantum field theory,
albeit a nonrelativistic one.  Quantization of electromagnetic waves produced the
first relativistic quantum field theory, which when supplemented by the quantized
Dirac field gave us quantum electrodynamics, whose further
generalization to matrices of fields is the present-day standard model of elementary
particles.  This development carries with it an extrapolation over enormous scales:
initial applications were at microscopic distances or at energies of a few electron
volts, while contemporary studies of elementary particles involve $10^{11}$~electron
volts or short distances of $10^{-16}$~cm.  The ``quantization" procedure, which
extended  classical field theory's  range of validity, consists of expanding a
classical field in normal modes, and taking each mode to be a quantal oscillator.

Field theoretic ideas also reach for the cosmos through the
development of the ``inflationary scenario''  --- a speculative, but completely physical
analysis of the early universe, which  appears to be consistent with
available observations.
Additionally, quantum field theories provide effective descriptions of many-body, 
condensed matter physics.  Here the excitations are not elementary particles and
fundamental interactions are not probed, but the collective phenomena that are
described by  
 many-body field theory exhibit
many interesting  effects, which in turn have been recognized as important for 
elementary particle theory.  Such exchanges of ideas between different
subfields of physics demonstrate  vividly the vitality and flexibility of field theory.

But in spite of these successes, today there is little confidence that field theory will
advance our understanding of Nature at its fundamental workings, beyond what has
been achieved.  While in principle all observed phenomena can be
explained by present-day field theory (in terms of the quantal standard model for
particle physics and the classical Newton--Einstein model for gravity),  these
accounts are still imperfect.  The particle physics model requires a list of {\it
ad hoc\/} inputs that give rise to conceptual, general questions such as:  Why is the
dimensionality of space-time four? Why are there two types of elementary particles
(bosons and fermions)? What determines the number of species of these particles? 
The standard model also leaves us with specific technical questions: 
What fixes the matrix structure, various mass parameters, mixing angles, and
coupling strengths that must be specified for concrete prediction?  Moreover,  classical
gravity theory has not been integrated into the quantum field description of
nongravitational forces, again because of conceptual and technical obstacles: 
quantum theory makes use of a fixed space-time, so it is unclear how to quantize
classical gravity, which allows space-time to
fluctuate; even if this is ignored, quantizing the metric tensor of Einstein's theory
produces a quantum field theory beset by infinities that cannot be controlled.

But these shortcomings are actually symptoms of a deeper lack of
understanding that has to do with symmetry and symmetry breaking.  Physicists 
mostly agree that ultimate laws of Nature enjoy a high degree of symmetry, that is,
the formulation of these laws is unchanged when various transformations are
performed.  Presence of symmetry implies absence of complicated and irrelevant
structure, and our conviction that this is fundamentally true reflects an ancient
aesthetic principle: physicists are happy in the belief that Nature in its
fundamental workings is essentially simple.  However, we must also recognize that
actual, observed physical phenomena rarely exhibit overwhelming regularity. 
Therefore, at the very same time that we construct a physical theory with intrinsic
symmetry, we must find a way to break the symmetry  in physical consequences of
the model.

Progress in physics can frequently be seen as the resolution of this tension.  In
classical physics, the principal mechanism for symmetry breaking is through
boundary and initial conditions on dynamical equations of motion.  For example,
Newton's rotationally symmetric gravitational equations admit the rotationally 
nonsymmetric solutions that describe actual orbits in the solar system, when
appropriate, rotationally nonsymmetric, initial conditions are posited.

The construction of physically successful quantum field theories makes use of
symmetry for yet another reason.  Quantum field theory models are
notoriously difficult to solve and also explicit calculations are beset by infinities. 
Thus far we have been able to overcome these two obstacles only when the models
possess a high degree of symmetry, which allows unraveling the complicated
dynamics and taming the infinities by renormalization.  Our present-day model for
quarks, leptons, and their interactions exemplifies this by enjoying a variety of chiral,
scale/conformal, and gauge symmetries.  But to agree with experiment, most of these
symmetries must be absent in the solutions.  At present we have available two
mechanisms for achieving this necessary result.  One is {\it spontaneous symmetry
breaking}, which relies on energy differences between symmetric and asymmetric
solutions:  the dynamics may be such that the asymmetric solution has lower energy
than the symmetric one, therefore the former is realized in Nature while the latter is
unstable.  The second is {\it anomalous\/} or {\it quantum mechanical symmetry
breaking}, which uses the infinities of quantum theory to effect a violation of the
correspondence principle:  the symmetries that appear in the model {\it before\/}
quantization disappear {\it after\/} quantization, because the renormalization
procedure --- needed to tame the infinities and well define the theory --- cannot be
carried out in a fashion that preserves the symmetries.

While these two methods of symmetry breaking successfully reduce the symmetries
of the standard  model to a phenomenologically acceptable level, this reduction is
achieved in an {\it ad hoc\/} manner, and much of the previously mentioned
arbitrariness, which must be fixed for physical prediction, arises precisely because of
the uncertainties in the symmetry-breaking mechanisms.  Spontaneous symmetry
breaking is adopted from many-body, condensed matter physics, where it is
well understood:  the dynamical basis for the instability of symmetric configurations
can be derived from first principles.  In the particle physics application, we have not
found the dynamical reason for the instability.  Rather, we have
postulated that additional fields exist, which are destabilizing and accomplish the
symmetry breaking.  But this {\it ad hoc\/} extension introduces additional, {\it a
priori\/} unknown parameters and yet-unseen particles, the Higgs mesons. 
Anomalous symmetry breaking also carries with it arbitrariness: the QCD
``renormalization scale" as well as yet-unseen particles, the axions, which fix an
arbitrary CP-violating angle.  Moreover, the field theoretic infinities, which give rise
to anomalous symmetry breaking, prevent the construction of an acceptable quantum
gravity field theory, so it is peculiar to rely on them so critically for the viability of
the standard model.  

Advancing our understanding of the above has been at an impasse for over
two decades.  In the absence of new experiments to channel theoretical speculation, 
some physicists have concluded that it will not be possible to make progress on these
questions within field theory, and have turned to a new structure, {\it string theory}. 
In field theory the quantized excitations are point particles with point interactions
and this gives rise to the infinities.  In string theory, the excitations are extended
objects --- strings --- with nonlocal interactions; there are no infinities, and this enormous
defect of field theory is absent.  Not only does quantum gravity exist in the new
context, but it appears that some puzzles having to do
with black holes can be answered.  Moreover, string theory addresses precisely some
of the questions that remained unanswered in field theory:  dimensionality of
space-time cannot be arbitrary because string theory cannot be formulated in
arbitrary dimensions; fermions must coexist with bosons because of supersymmetry --- a necessary ingredient of string theory; and~so~on.

Yet in spite of these positive features, up to now string theory provides only a
framework, rather than a definite structure.  While present-day physics should be
found in the low-energy limit of string theory, a precise derivation of the standard
model has yet to be given.  One thinks again about symmetry and
symmetry breaking.  The symmetries of quantum field theory surpass those of
classical physics and require elaborate symmetry breaking mechanisms.  The
symmetries of string theory again vastly outpace those of field theory, and must be
broken by  yet-to-be-developed procedures, in order to explain the world around
us.

On previous occasions when it appeared that quantum field theory was incapable of
advancing our understanding of fundamental physics, new approaches and new ideas
dispelled the pessimism.  Today we do not know whether our lack of progress is due
to a failure of imagination, or whether indeed we have to present fundamental
physical laws in a new framework,  thereby
replacing the field theoretic one, which has served us well for over a hundred years.

\end{document}